\documentstyle[12pt,amssymb,epsfig]{article}
\begin{document}
\begin{center}
\begin{minipage}{14.5cm}
\begin{flushright}
{\bf LAL 02-78}\\

November 2002
\end{flushright}
\end{minipage}
\end{center}
\vspace{3.2cm}
\centerline {\LARGE\bf Structure function results from H1}
\renewcommand{\thefootnote}{\alph{footnote}}
\vspace{1.5cm}
\centerline {{\sc {\sc Zhiqing Zhang}\footnote{zhangzq@lal.in2p3.fr}}}
\centerline {On behalf of the H1 Collaboration}
\vspace{0.5cm}
\centerline {\it Laboratoire de l'Acc\'el\'erateur Lin\'eaire,
 IN2P3-CNRS et}
\centerline {\it Universit\'e de Paris-Sud, BP 34, 
 91898 Orsay Cedex}
\vspace{3cm}
\centerline {Abstract}
\bigskip
\begin{center}
\begin{minipage}{14.5cm}
{New structure function results from H1 are presented. The measurements
cover a huge kinematical range for $Q^2$, the four momentum transfer
squared, from 0.35\,GeV$^2$ to 30\,000\,GeV$^2$, and for Bjorken $x$
between $\sim 5\cdot 10^{-6}$  and 0.65.
At $Q^2>100$\,GeV$^2$, full HERA\,I data
have been analyzed. The data are compared with a new QCD analysis. 
The impact of the HERA\,I data on the parton density functions is discussed.}
\end{minipage}
\end{center}
\newpage
\renewcommand{\thefootnote}{\arabic{footnote}}
\setcounter{footnote}{0}

\section{Introduction}

Deep inelastic scattering (DIS) experiments have played an important role 
in the understanding of the partonic structure of matter and 
in establishing the strong interaction sector of the Standard Model (SM): QCD.

In the past decade, the HERA experiments H1 and ZEUS have been providing 
more and more precise data for modern global fits allowing parton density 
functions (PDFs) be determined with steadily increasing precision. 
The precision of the PDFs is of importance as it is needed to have reliable 
predictions for precision measurements and for new physics searches at 
future hadron colliders such as LHC.

In this talk, I shall first present the latest new structure function 
measurements from H1, then try to answer two following questions: 
i) Can HERA\,I cross section or structure function (SF) data measured so far 
be described by QCD? and ii) What is the impact of the HERA\,I data? 

\section{New Measurements and Impact of HERA\,I Data}

Four new preliminary measurements\,\cite{h1-f2-ichep02} from H1 
have been submitted to this conference. Thanks to the increased proton beam
energy from 820\,GeV to 920\,GeV since 1998 (the center-of-mass energy 
$\sqrt{s}$ is accordingly increased from 300\,GeV to 320\,GeV), the fully
installed backward silicon tracker, the new capability of triggering
low energy scattered electrons\footnote{It refers generically to
both electrons and positrons unless stated otherwise.} down to 3\,GeV, 
and the special runs where data were taken by shifting the nominal
interaction vertex in the proton beam direction by 70\,cm,
the kinematical domain has been substantially extended with respect to 
previous measurements both to lower
$Q^2$ and to smaller $x$.
On the other extreme at high $Q^2$ and large $x$, the new
measurements of both neutral current (NC) and charged current (CC)
processes are obtained with improved precision using the highest
data sample taken before the HERA machine was upgraded for the
second phase.

\begin{figure}[htb]
\begin{center}
\begin{picture}(50,280)
\put(-150,-15){\epsfig{file=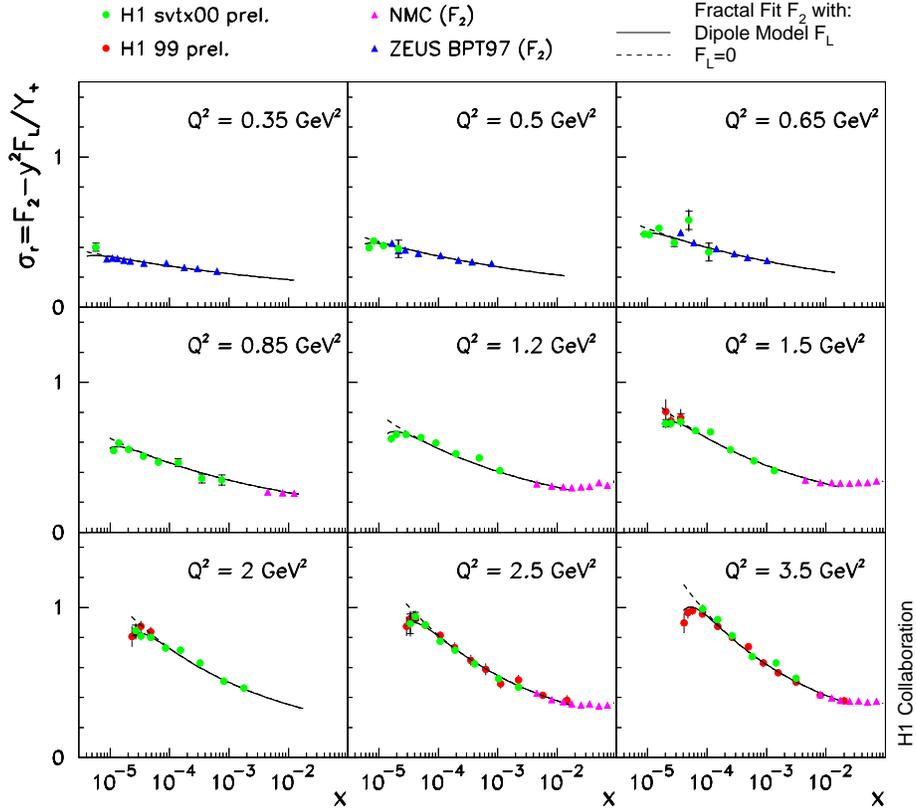,width=12cm}}
\end{picture}
\end{center}
\caption{Preliminary reduced cross section (see text) measurements from H1
shown as a function of $x$ for $Q^2\leq 3.5$\,GeV$^2$. Also shown
are measurements from NMC\,\cite{nmc} and ZEUS\,\cite{zeus-bpt} and
expectations based on a phenomenological model\,\cite{fractal-fit}.}
\label{fig:rxs-lowq2}
\end{figure}

Part of the new measurements at low $Q^2$, the so-called reduced 
cross sections $\tilde{\sigma}$ (or $\sigma_{\rm r}$), 
are shown in
Fig.\,\ref{fig:rxs-lowq2}. 
The reduced cross section for the NC
interaction may be expressed in terms of SFs as
\begin{equation}
\tilde{\sigma}_{\rm NC}(e^\pm p)=F_2-\frac{y^2}{Y_+}F_L\mp\frac{Y_-}{Y_+}xF_3
\end{equation}
where $Y_\pm=1\pm (1-y)^2$ with $y=Q^2/(xs)$,  
$F_2$ is the dominant contribution, 
the contribution of the longitudinal SF $F_L$ is suppressed at low $y$ by 
$y^2/Y_+$, and the SF $xF_3$, which arises mainly from the photon-$Z^0$
interference and has been previously determined\,\cite{hera-xf3} at 
$Q^2\geq 1500$\,GeV$^2$, becomes non-negligible only at such large $Q^2$ 
values.
The reduced cross section differs from the experimentally measured double 
differential cross section ${\rm d}^2\sigma/{\rm d}x{\rm d}Q^2$ by 
a kinematical factor of $2\pi\alpha^2Y_+/xQ^4$, where $\alpha$ is the fine 
structure constant.

From Fig.\,\ref{fig:rxs-lowq2}, one observes that the new measurements
extend now down to $x\sim 5\cdot10^{-6}$ at $Q^2=0.35$\,GeV$^2$. The rise
of $F_2$ as $x$ decreases, the first important observation at
HERA and being related to increasing gluon density $xg$ at small $x$,
persists even at $Q^2<1$\,GeV$^2$ although with
a reduced magnitude.
%, showing no sign yet of $xg$ saturation at low $x$. 
The same observation is also obtained from the
measurement of ${\rm d}\!\ln F_2/{\rm d}\!\ln x$ (not shown here).

\begin{figure}[htb]
\begin{center}
\begin{picture}(50,375)
\put(-150,-15){\epsfig{file=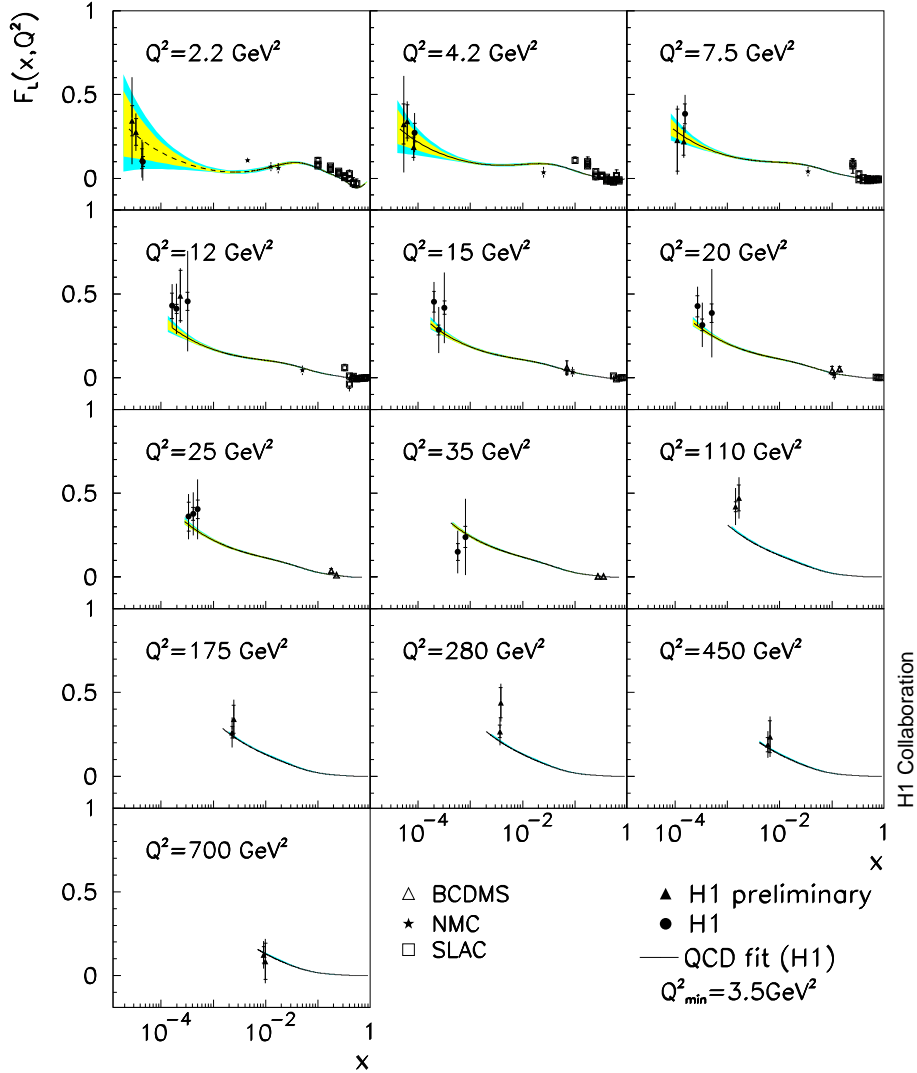,width=12cm}}
\end{picture}
\end{center}
\caption{Preliminary determination of $F_L$ shown together with
those by H1\,\cite{h1-oldfl} and BCDMS, NMC, and SLAC\,\cite{ft-fl}. 
Also shown is the SM expectation - QCD fit (H1)\,\cite{h1-oldfl}.}
\label{fig:fl}
\end{figure}

As $Q^2$ increases, the measured $\sigma_{\rm r}$ is observed to deviate
from the dashed curves at high $y$ in which the $F_L$ is set to zero,
but agree with the full curves where a non-zero $F_L$ according to a
dipole model is included. Such a sensitivity to $F_L$ has been
exploited by H1 to determine experimentally $F_L$, which is shown in
Fig.\,\ref{fig:fl}. The new determination (shown in triangle) thus
extends the previous determination\,\cite{h1-oldfl} to smaller $x$ and
higher $Q^2$. It should be pointed out that such a determination,
though not a direct measurement which needs different beam energies,
does provide a non-trivial consistency test as the determination gives
a direct measure of $xg$ while the expectation from a QCD fit
was based on an indirect $xg$ derived from the scaling violation of
the $F_2$.

\begin{figure}[htb]
\begin{center}
\begin{picture}(50,265)
\put(-150,-35){\epsfig{file=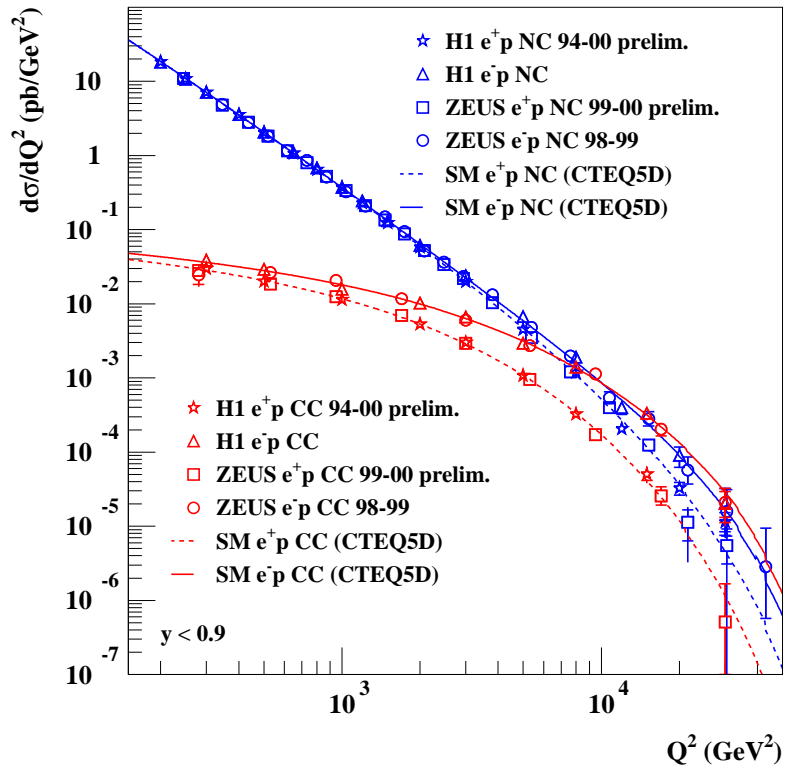,width=12cm}}
\end{picture}
\end{center}
\caption{NC and CC cross sections ${\rm d}\sigma/{\rm d}Q^2$ measured with
$e^+p$ and $e^-p$ data by H1\,\cite{h1-f2-ichep02,hera-xf3} and
ZEUS\,\cite{zeus-9900,hera-xf3} in comparison with the SM
expectations from CTEQ5\,\cite{cteq5}.}
\label{fig:nccc-q2}
\end{figure}

Moving to even higher $Q^2$ (100--30\,000\,GeV$^2$) region,
new preliminary cross section measurements
based on the $e^+p$ 1999-2000 at $\sqrt{s}=320$\,GeV\,\cite{h1-f2-ichep02} 
are combined with those measured with the $e^+p$ 1994-1997 at 
320\,GeV\,\cite{e+9497} taking into account a small correction due to 
the increased $\sqrt{s}$.
The combined cross sections in $e^+p$ scattering can be compared with those
in $e^-p$ scattering\,\cite{hera-xf3}.
In Fig.\,\ref{fig:nccc-q2}, the single differential NC and CC cross sections
from H1\,\cite{h1-f2-ichep02,hera-xf3} and ZEUS\,\cite{zeus-9900,hera-xf3}
are shown and compared with the SM expectation from a global 
QCD fit (CTEQ5\,\cite{cteq5}).
While the NC $e^+p$ and $e^-p$ cross sections are indistinguishable at
low $Q^2$ as expected from the photon ($\gamma$) exchange, 
they are different at high $Q^2$
due to the positive (negative) $\gamma Z^0$ exchange in $e^-p (e^+p)$
interactions.
The CC $e^+p$ and $e^-p$ cross sections are different over the entire
$Q^2$ region measured originating from different quarks probed by negatively 
and positively charged W bosons exchanged in the interactions and from
different helicity structures involved.
At $Q^2\sim M^2_{Z,W}$, all four cross sections (NC, CC and $e^+p$,
$e^-p$) are comparable demonstrating the
universal coupling strength of electroweak interactions.
It is also remarkable that the cross sections which vary over many orders of 
magnitude are described by the global fit which to some extent is independent
of the data.

HERA data have been important for constraining the gluon density $xg$ at
small $x$. Indeed, in an earlier QCD analysis performed by H1
using the H1 low $Q^2$ precision data and part of the H1 high $Q^2$ $e^+p$
data available then, the gluon density $xg$ was determined with
an experimental precision of about 3\%\,\cite{h1-oldfl}.
In addition, the strong coupling constant $\alpha_s$ was determined with
an experimental uncertainty of 1.5\%\,\cite{h1-oldfl} using the same data 
together with the proton data from BCDMS\,\cite{bcdms} for constraining $xg$
at large $x$.
This determination at next-to-leading order (NLO) is unfortunately limited 
theoretically by the large scale uncertainty showing the necessity
of having next-to-NLO calculations for improvements.

\begin{figure}[htb]
\begin{center}
\begin{picture}(50,270)
\put(-150,-30){\epsfig{file=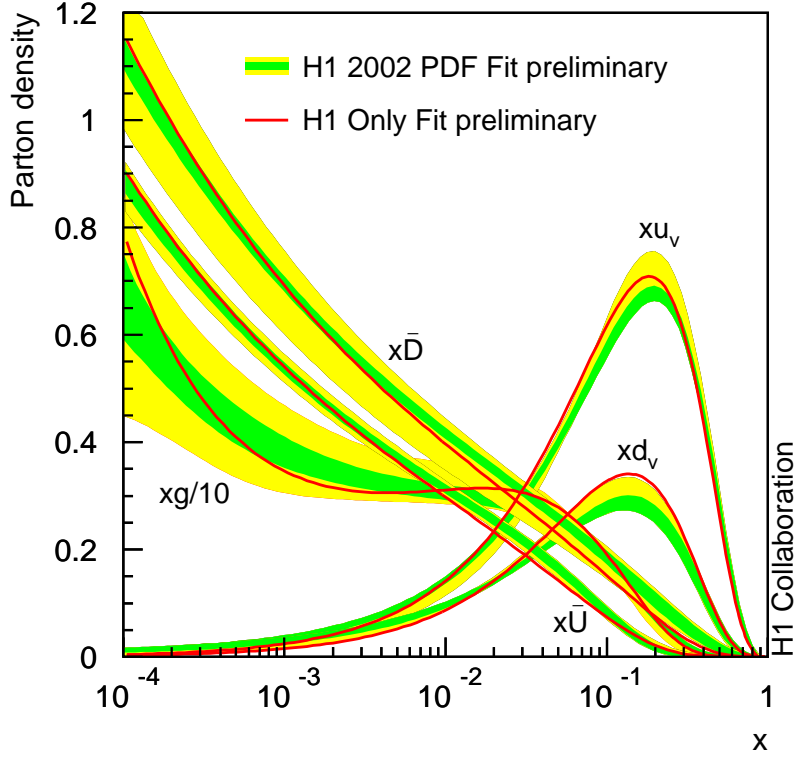,width=12cm}}
\end{picture}
\end{center}
\caption{The valence quark densities $xu_v, xd_v$, the anti-up and down
type densities $x\overline{U}, x\overline{D}$ and the gluon density $xg$
(reduced by a factor of 10), from the combined fit (error bands) 
using the H1 and BCDMS data\,\cite{bcdms} are compared with those from 
the fit (solid curves) using the H1 data only.}
\label{fig:h1pdf02}
\end{figure}

Presented at this conference is a further NLO QCD analysis performed by H1
using the low $Q^2$ precision data and all high $Q^2$ cross section data.
%which is similar to those global analyses from
%CTEQ\,\cite{cteq5} and MRST\,\cite{mrst}.
%In order to see the impact of HERA\,I data, the analysis is based on the H1
%data only or on a minimum data sets from fixed target experiments in
%addition.
The analysis aims at a determination of the quark momentum distributions, 
besides $xg$.
A novel feature of the analysis is the way in which the various structure
functions are decomposed in terms of different parton densities.
The quark densities used by H1, the upper and down types of quarks
($xU$, $xD$) and their anti-quarks ($x\overline{U}$, $x\overline{D}$),
directly enter the NC and CC cross sections contrary to 
an effective sea distribution or valence quark distributions. 
The latter follow from the differences $u_v=U-\overline{U}$ and 
$d_v=D-\overline{D}$. 
The initial distributions are parametrized in an MRST-like functional form 
for $Q^2_0$ at 4\,GeV$^2$.
The results of the fit are shown in Fig.\,\ref{fig:h1pdf02} as a function of 
$x$.
In the fit with the full HERA\,I data plus the BCDMS proton and deuterium 
data\,\cite{bcdms},
the experimental uncertainty shown in dark error band reaches a precision of
a few percent. Shown in light error band is the model uncertainty,
which includes variation of $Q_0^2$ and other parameters in the fits.
The results of the fit using the H1 data only is shown with the solid curves.
It is remarkable that the fit using HERA data only is able to give
consistent results on the parton density functions with the combined fit which
uses additional data at low $Q^2$ and large $x$.

\begin{figure}[htb]
\begin{center}
\begin{picture}(50,205)
\put(-165,-250){\epsfig{file=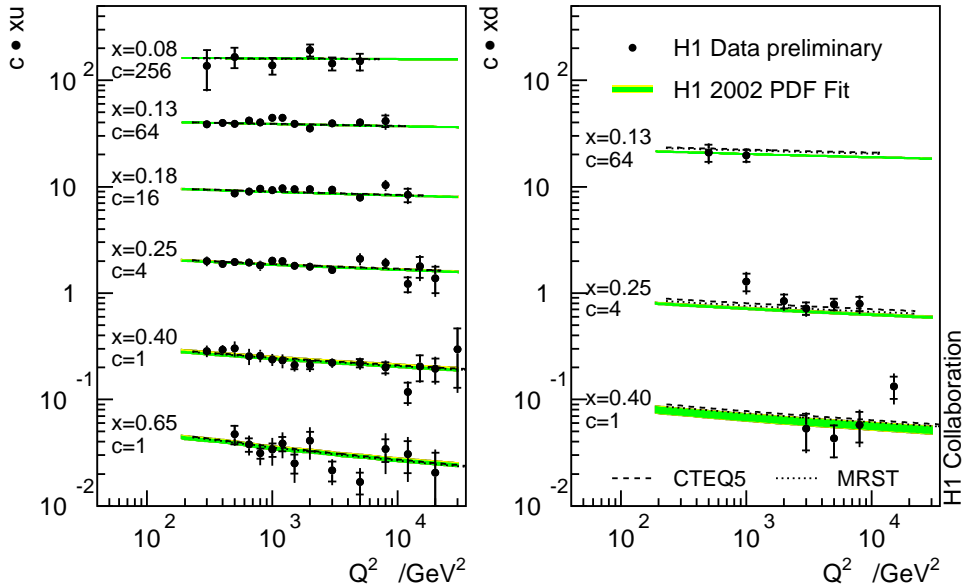,width=14cm}}
\end{picture}
\end{center}
\caption{The $u$ and $d$ quark densities determined from the local
extraction method (full points) and from the fit are compared with
those from CTEQ5\,\cite{cteq5} and MRST\,\cite{mrst}. For displaying
purpose, the densities are multiplied by a factor $c$.}
\label{fig:ud}
\end{figure}

The impact of the HERA\,I data at large $x$ is best seen on the u and d quark
densities at valence quark region, determined by two complementary
methods (Fig.\,\ref{fig:ud}).
In one method, the densities are determined from the fit, where the precision
is optimal, while in the second method, the densities ($xq_{\rm exp}$)
are extracted locally from the measured cross sections 
$\tilde{\sigma}_{\rm meas}(x,Q^2)$, $xq_{\rm
exp}=\tilde{\sigma}_{\rm meas}[xq/\tilde{\sigma}]_{\rm fit}$. 
The latter method has the advantage
that the resulting $u$ and $d$ densities do not depend on the $Q^2$ evolution 
and are thus free from any nuclear corrections.

\section{Conclusion}

To conclude, the structure function data from almost a decade's operation 
in HERA phase I have enabled the theory of QCD be tested to a new
level of precision in deep inelastic scattering and 
revealed its impact on the parton densities both
for the gluon at small $x$ and for the $u$ and $d$ at large $x$. The
full HERA\,I data at large $x$ and high $Q^2$ are still statistically
limited. New data to be taken during the next years in HERA phase
II are therefore important not only to improve the precision of the
structure functions but also to fully explore the very high $Q^2$
region up to the kinematical limit $s\sim 10^5$\,GeV$^2$.

\end{document}